\documentclass[12pt]{iopart}
\usepackage[utf8]{inputenc}
\usepackage[T1]{fontenc}
\usepackage{ae,aecompl} 
\usepackage{graphicx}
\graphicspath{{figures/}} % Directory in which figures are stored
\usepackage{color}
\usepackage{amssymb}
\usepackage{latexsym}
\usepackage{wasysym}
\usepackage{psfrag}
\usepackage{ifthen}
\usepackage[citecolor=blue,colorlinks=true]{hyperref}
\usepackage{longtable}
\usepackage{float}
\usepackage[utf8]{inputenc}
\usepackage{lineno}
\usepackage{units}
\usepackage{multirow}
\usepackage{ulem}

\newcommand{\changes}[1]{{\color{black}#1}}
\newcommand{\newchanges}[1]{{\color{black}#1}}
\newcommand{\revisions}[1]{{\color{black}#1}}

\begin{document}

\title[Simulations of light distribution on new Virgo instrumented baffles]{Simulations of light distribution on new instrumented baffles surrounding Virgo end mirrors}

\author{A. Macquet$^1$, M. Andrés-Carcasona$^1$, M. Mart\'inez$^{1,2}$, Ll.M. Mir$^1$, A. Romero-Rodr\'{\i}guez$^{1}$, and H. Yamamoto$^4$}

\address{$^1$Institut de F\'isica d’Altes Energies (IFAE), The Barcelona Institute of Science and
Technology, Campus UAB, 08193 Bellaterra (Barcelona) Spain}
\address{$^2$Instituci\'o Catalana de Recerca i Estudis Avancats (ICREA), Barcelona, Spain}
% \address{$^3$Theoretische Natuurkunde, Vrije Universiteit Brussel, Pleinlaan 2, B-1050 Brussels, Belgium}
\address{$^4$LIGO laboratory, California Institute of Technology (Caltech), Pasadena, CA, US}

\vspace{10pt}

\begin{abstract}
As part of the second phase of Advanced Virgo upgrade program, instrumented baffles are being constructed to be installed around the end mirrors in the main arms, in continuation of what has been implemented for the input mode cleaner end mirror during phase I. These baffles will be equipped with photosensors, allowing for real-time monitoring of the stray light around the mirrors. In this paper, we present optical simulations of the light distribution in the detector's main cavities to assess the ability of the sensors to effectively monitor misalignment and defects on the mirrors' surface and to play a role in the pre-alignment of the interferometer.
% The effect of the backscattered light from the baffles is also computed and projected over the O5 sensitivity curve,  to evaluate possible effects of the presence of instrumented baffles on the ultimate sensitivity of the detector.
\end{abstract}

%\maketitle

\section{Introduction}

% Generic intro - Adv Virgo upgrade program
In its final configuration, Advanced Virgo \changes{Plus} will be a dual-recycled Fabry-Perot-Michelson interferometer with a detection range \changes{between \unit[$150-260$]{Mpc}} for gravitational waves emitted by binary neutron 
stars mergers~\cite{10.1117/12.2565418}. To achieve this level of sensitivity, two phases of upgrades are planned with several improvements intended to reduce the level of noise in the detector. The first phase is currently ongoing, and the second one will take place between the O4 and O5 observing runs, in 2024-2025.

\begin{figure}[htp]
    \centering
    \includegraphics[width=0.7\textwidth]{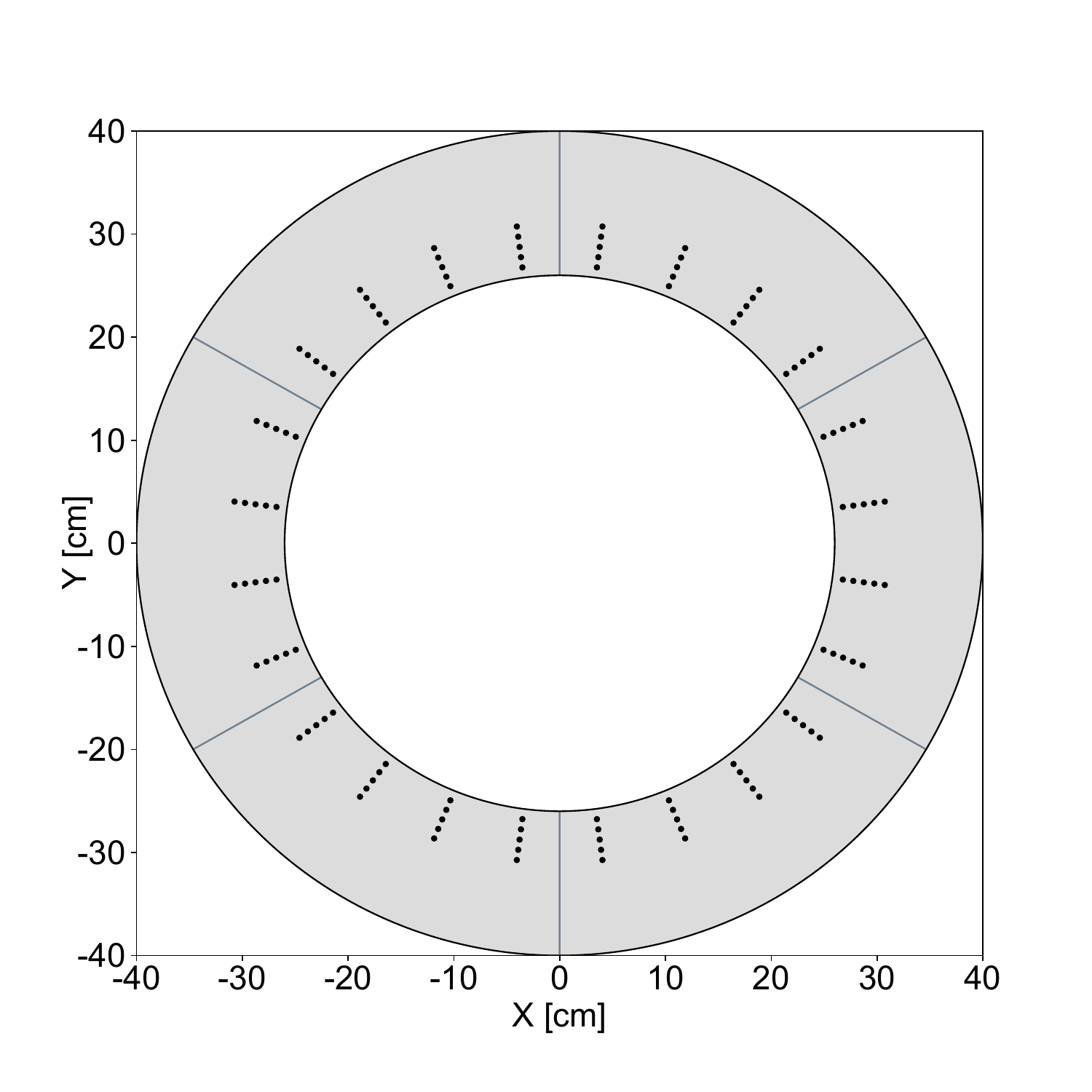}
    \caption{Planned layout of the sensors on the EM baffle.}
    \label{fig:layout}
\end{figure}

% Instrumented baffles
As part of this upgrade program, new instrumented baffles will be installed around the end mirrors (EM) in the main arms, in order to provide an active monitoring of the distribution of scattered light at low angles. \newchanges{It is also planed to instrument the baffles surrounding the input mirrors (IM) after O5.}
A first instrumented baffle was installed in April 2021 around the EM of the input mode cleaner (IMC) to demonstrate this new technology.
First measurements have been presented in Ref.~\cite{Ballester_2022}, and are in good agreement with previous simulations of stray light in the IMC cavity~\cite{2021CQGra..38d5002R}.
The baffles in the main arms will be instrumented with $120$ Si-based photodiodes provided by Hamamatsu \revisions{photonics} in Japan. They are
similar to the ones installed on the IMC baffle~\cite{Ballester_2022, IMClongPaper}, \revisions{based on the S13955-01 model.} \newchanges{These sensors will have a dynamic range of \unit[$13$]{mW} and a resolution of approximately \unit[$60$]{$\mu$W}. They will be read at a rate up to \unit[$1$]{kHz}, allowing to correlate with interferometer glitches.} 
\newchanges{The inner and outer radii of the baffle will be \unit[$26$]{cm} and \unit[$40$]{cm}, respectively, }and the sensors will be arranged in five concentric rings at \unit[$27$, $28$, $29$, $30$, and $31$]{cm} from the center of the EM. \newchanges{No sensor will be placed beyond $31$ cm, because the field is shielded by a baffle located in the cryotrap at $5.4$ m from the EM, which has an aperture of $60$ cm.} In the planned configuration, each ring will be populated by $24$ photodiodes. The layout is shown in Figure~\ref{fig:layout}, although the final design and total number of sensors may be subject to minor modifications due to technical constraints. 

In this paper, we assess the merits of these instrumented baffles using optical simulations of the light distribution in the main arms. First, we describe the simulation tool and the parameters used for this study. Then, we report the distribution of surface power on and around the main mirrors to estimate the amount of light that will be received by the baffles. We compute the power that will be seen by each sensor in different configurations, introducing misalignment and point defects on the mirrors, in order to assess the ability of the sensors to detect such defects. 
% Finally, we model the backscaterring from the baffle and sensors to evaluate their potential effect on the ultimate sensitivity of the detector.

\section{Principle of the simulations}
% Description of SIS principle and the way simulations are done
We use the \textit{Stationary Interferometer Simulation}, SIS~\cite{SIS20}, to compute the fields in the main arms. SIS relies on fast Fourier transforms to calculate the propagation of the field in a realistic optical cavity, taking into account the roughness of mirrors, coating absorption and thermal deformations. The 2D distribution of the field can be extracted at any point of the cavity using an adaptative grid.

\begin{figure}[htp]
    \centering
    \includegraphics[width=0.8\textwidth]{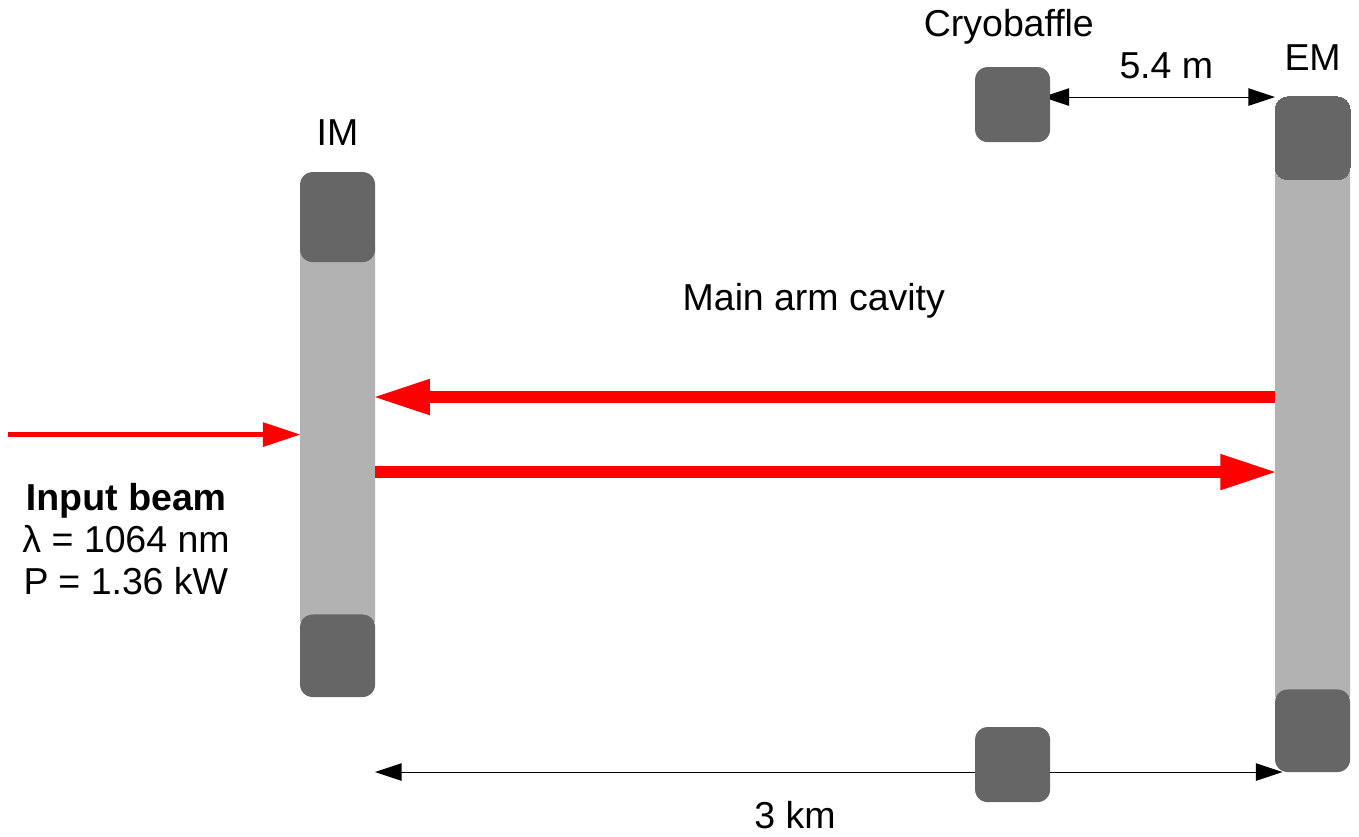}
    \caption{Sketch of the optical system simulated that represents a single Virgo arm. The power of the beam impinging on the IM is \unit[$1.36$]{kW}, which corresponds to the expected value for O5 after the power recycling cavity. Baffles are represented in dark grey, and mirrors in light grey.}
    \label{fig:sketch}
\end{figure}

A sketch of the optical setup simulated is shown in Figure~\ref{fig:sketch}. It consists in a \unit[$3$]{km}-long Fabry-Perot cavity that represents a Virgo main arm. 
\newchanges{
 To simulate the propagation of scattered light in the cavity, realistic maps of the mirrors' surface are used. For the IM, we use a surface map that has been measured during O3. Since the EM will be changed between O4 and O5, we do not have access to a measured surface map, so we use a simulated one~\cite{degallaix:hal-02417232}.} 
In addition, the effect of the baffle that surrounds the cryotrap \changes{(often called cryobaffle)} is taken into account by clipping the field at an aperture of \unit[$60$]{cm} at this location. The clipped field is then allowed to propagate freely to the EM. \changes{Another cryobaffle is present in front of the IM, but it is not modelled in these simulations as its aperture is larger than the outer diameter of the IM baffle.}

The main parameters used to perform the simulations are listed in Table~\ref{tab:input_nominal}. They correspond to the expected configuration of the detector during O5. The input power is \unit[$1.36$]{kW} and  corresponds to the \newchanges{maximal} expected power reaching the IM after the power-recycling cavity.
The \newchanges{resolution of the grid over which the field is computed at the level of the EM is \unit[$0.64$]{mm}.} 

\begin{table}[htp]
    \centering
    \begin{tabular}{l|l}
    \hline
    \multicolumn{2}{c}{Simulation parameters} \\
    \hline
    Input power & \unit[$1.36$]{kW} \\
    Wavelength & \unit[$1064$]{nm} \\

    EM aperture & \unit[$52$]{cm} \\
    IM aperture & \unit[$33$]{cm} \\
    EM transmitivity & $5 \cdot 10^{-6}$ \\
    IM transmitivity & $1.375 \cdot 10^{-2}$ \\
    EM RoC & \unit[$1969$]{m} \\
    IM RoC & \unit[$1067$]{m} \\

    EM thickness & \unit[$20$]{cm} \\
    IM thiskness & \unit[$20$]{cm} \\
%    \hline
%    \multicolumn{2}{c}{Simulation output} \\
%    \hline
%    Cavity power     &  \unit[$3.93 \cdot 10^5$]{W}\\
%    Power in EM baffle     & \unit[$2.2 \cdot 10^{-1}$]{W} ($0.6$ ppm) \\
%    Power in IM baffle     & \unit[$4.5 \cdot 10^{-1}$]{W} ($1.1$ ppm) \\
%    Beam width on EM     & \unit[$9.1$]{cm} \\
%    Beam width on IM     & \unit[$4.9$]{cm} \\

    \end{tabular}
%    \caption{Parameters used to simulate Virgo main arms during O5 and main output of the simulation.}
    \caption{Parameters used to simulate Virgo main arms during O5.}
    \label{tab:input_nominal}
\end{table}

We consider a layout of $5$ rings of $24$ sensors located at \unit[$27$, $28$, $29$, $30$, and $31$]{cm} from the center of the EM, as displayed in Figure~\ref{fig:layout}.
We note $P(\phi, r)$ the power received by a sensor located at angle $\phi$ and distance $r$ from the center of the mirror. This value is computed by integrating the surface power distribution over the area covered by the sensor. \newchanges{Since they will be placed behind conical holes of \unit[$4$]{mm} of diameter, their effective area is \unit[$0.13$]{cm$^2$}}. With a grid resolution of \unit[$0.64$]{mm}, several pixels cover the area of one sensor. 
We can then compute the amount of power received by each sensor in different configurations,
which will be shown at the end of next section.

\section{Results}
\subsection{Nominal configuration}
\begin{figure}[htp]
    \centering
    \includegraphics[width=0.45\columnwidth]{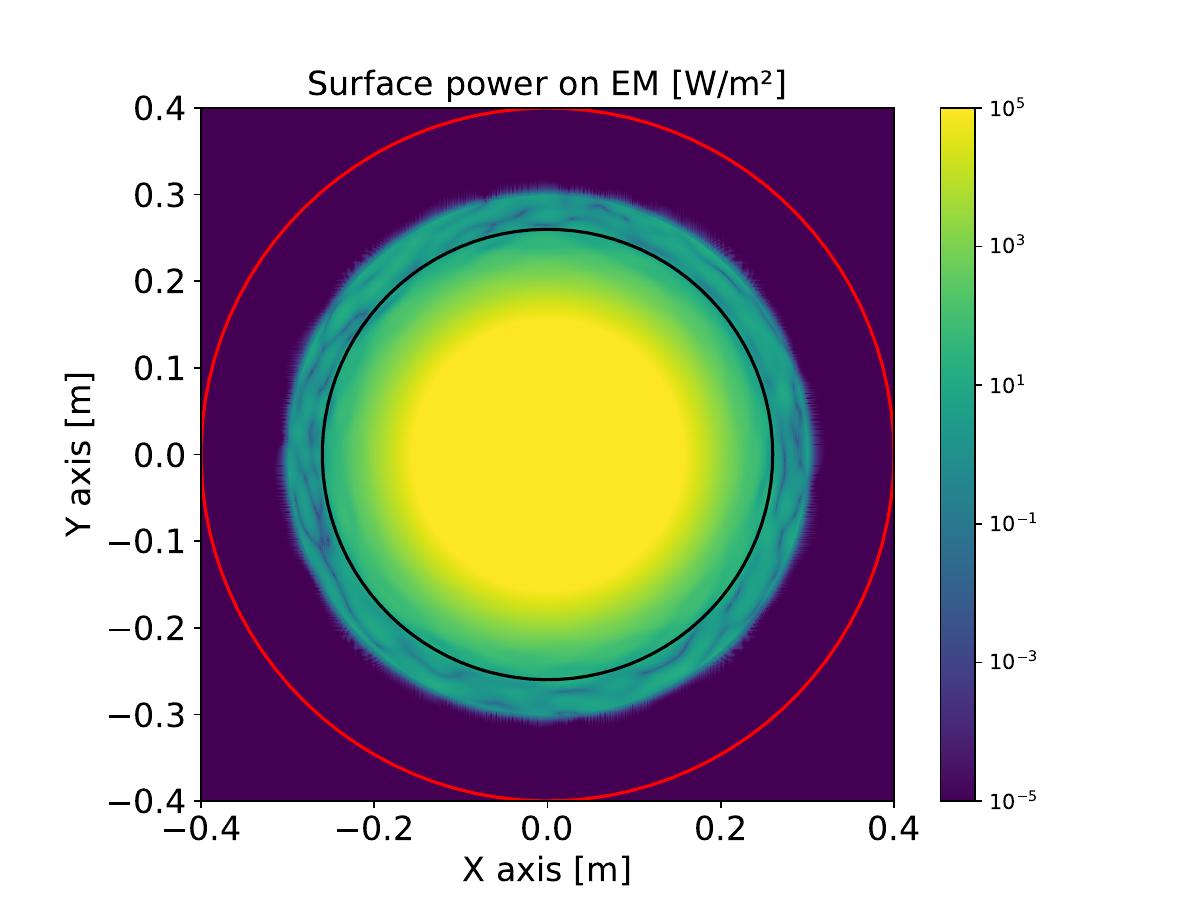}
    \includegraphics[width=0.45\columnwidth]{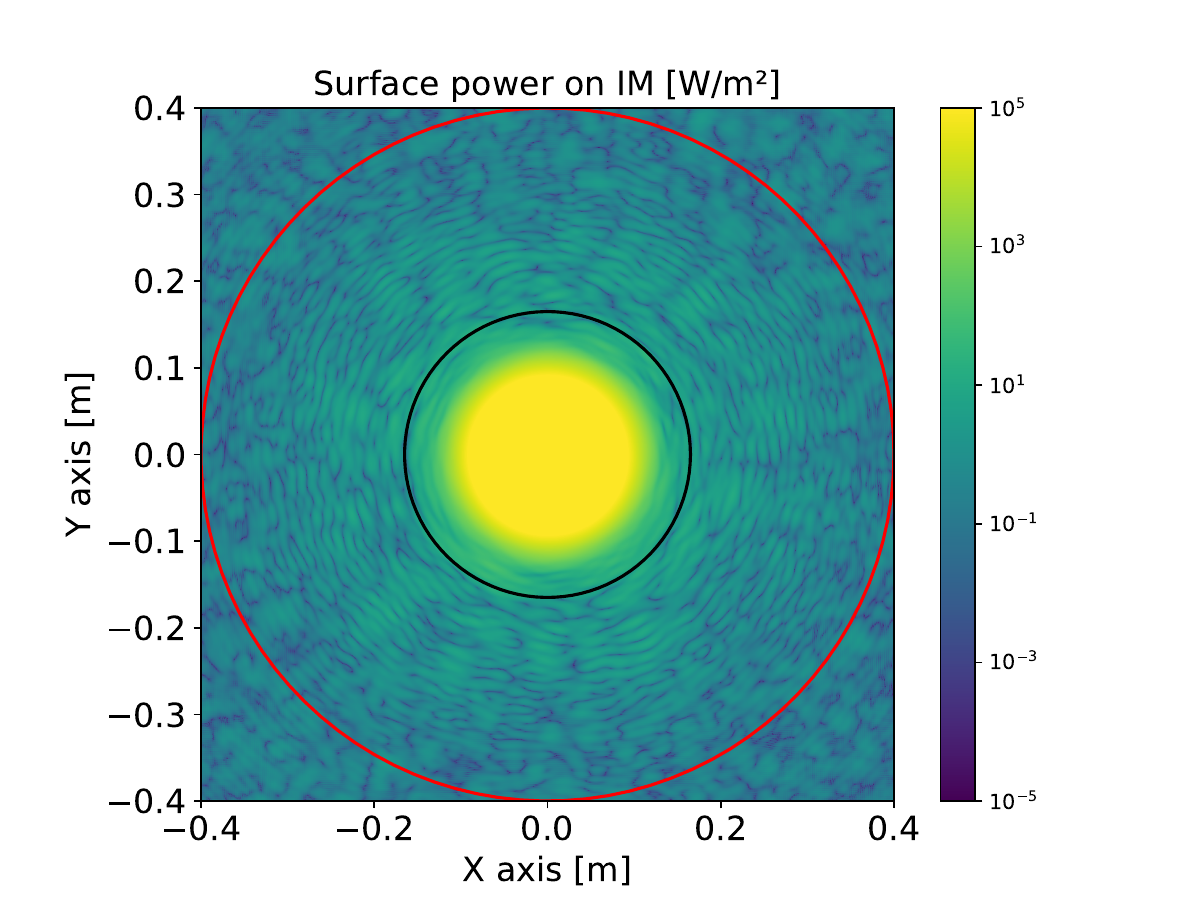}

    \caption{2-D maps of the surface power distribution on the EM (left) and IM (right) in nominal configuration (perfectly aligned cavity and realistic mirror maps). The black and red circles indicate the inner and outer radius of the baffle, respectively.}
    %White dots represent the sensors on the baffle surrounding the EM.}
    \label{fig:2D_nominal}
\end{figure}

% Results: field power distribution on IM and EM (2d maps), beam shape
We define the nominal configuration as the case where the cavity is perfectly aligned, and realistic mirror maps are used. The distribution of surface power in the ensemble mirror plus baffle is shown in Figure~\ref{fig:2D_nominal} for the EM and IM. The effect of the cryobaffle is clearly visible on the EM with a sharp cut-off of the surface power at a radius of \unit[$30$]{cm}.

\begin{figure}[htp]
    \centering
    \includegraphics[width=0.45\columnwidth]{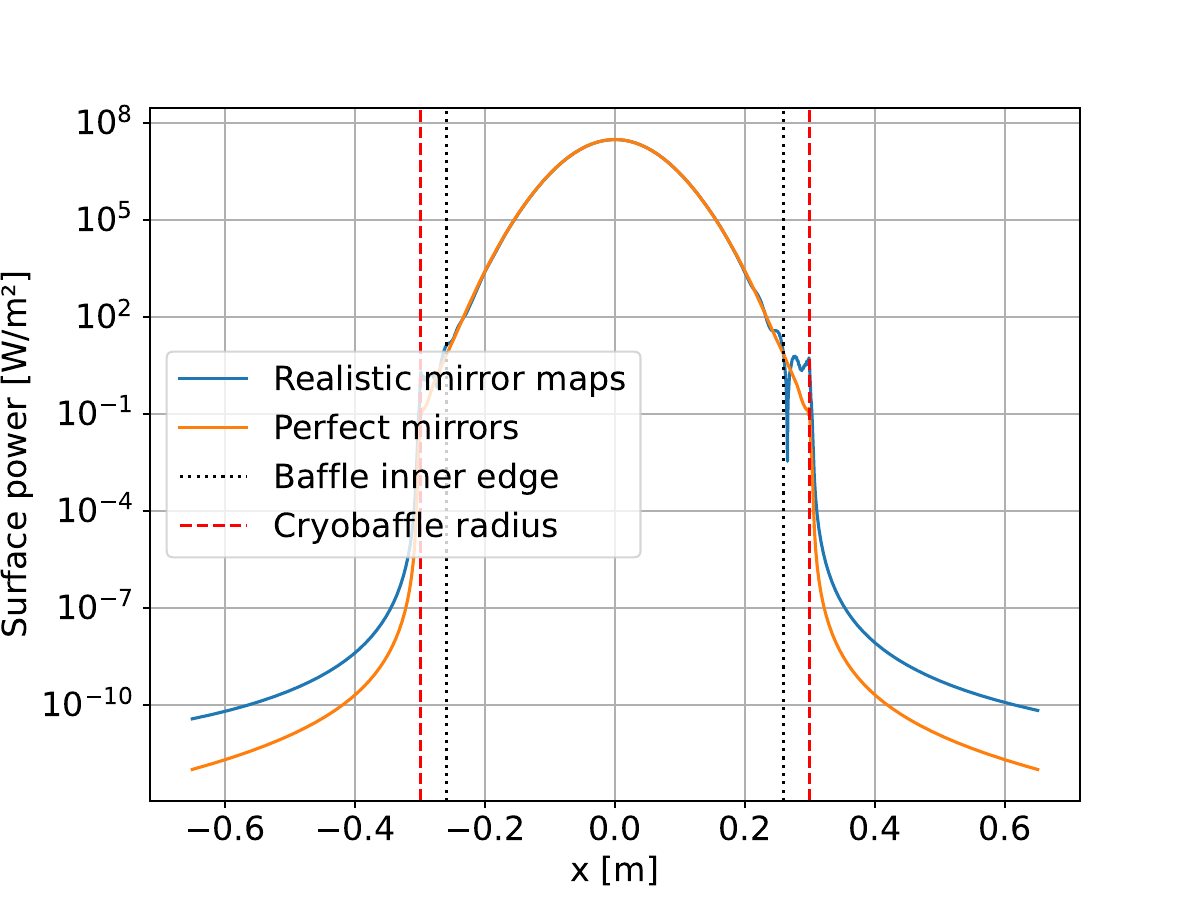}
    \includegraphics[width=0.45\columnwidth]{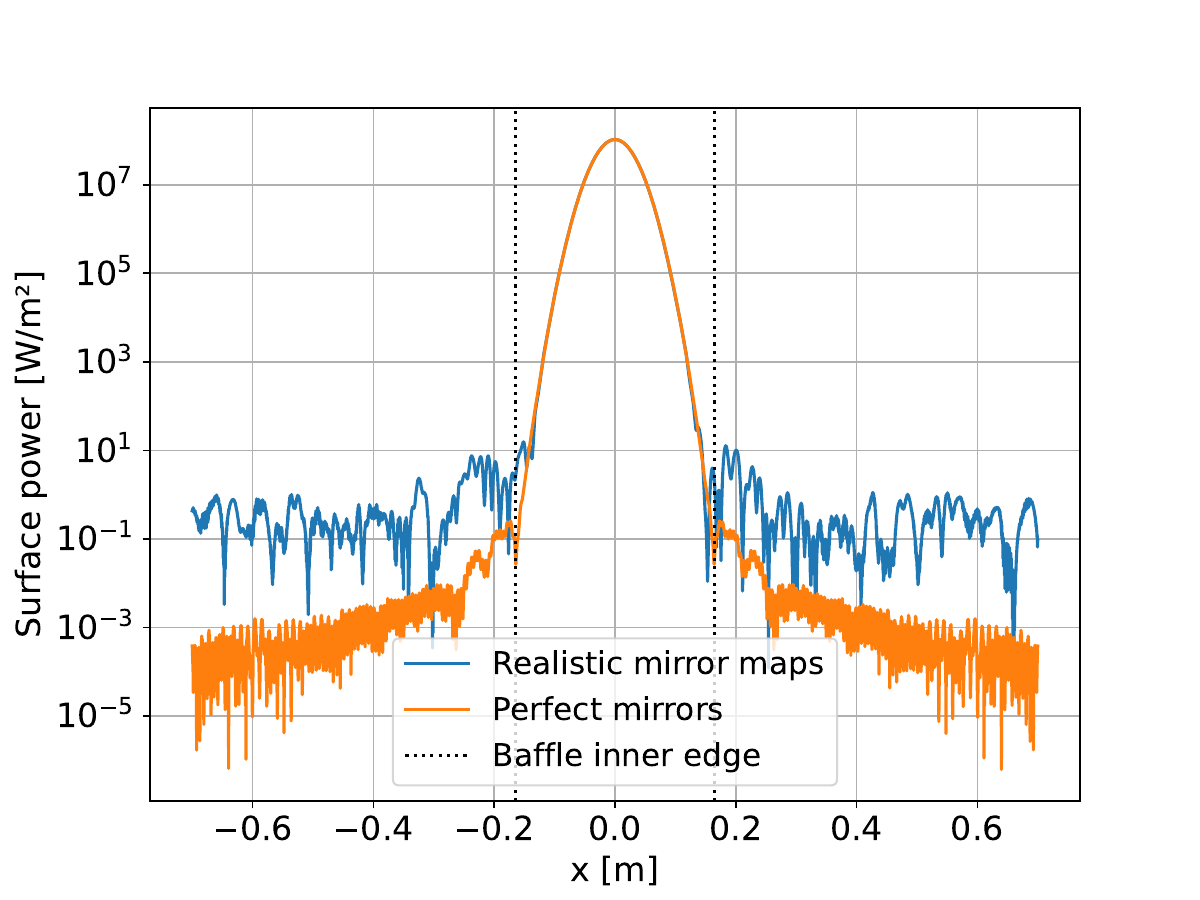}

    \caption{Distribution of power in EM (left) and IM (right) along the $x$ axis in nominal configuration with realistic mirror maps (blue) and with perfect mirrors (orange).
    The black vertical lines indicate the baffle inner edge,
    and the red vertical lines in the left figure indicate the inner radius of 
    the cryobaffle.}
    \label{fig:beam_nominal}
\end{figure}

The shape of the laser beam on the mirrors is represented in Figure~\ref{fig:beam_nominal}, in the case where realistic mirror maps are used, and in the case where perfect mirrors are assumed. In the central region, the beams are almost perfectly Gaussian with a width of \unit[$9.1$]{cm} and \unit[$4.9$]{cm} on the EM and IM, respectively. These values are consistent with expectations for O5~\cite{2020SPIE11445E..11F}.
The effect of mirror maps can be seen in the tails of the beam, which contain more power than with perfect mirrors. Thus, sensors located at radii between $27$ and \unit[$31$]{cm} are strategically placed to monitor scattered light. 

Table~\ref{tab:output_nominal} %\changes{also} 
shows the total power circulating in the cavity and reaching the baffles. The total power circulating is \unit[$393$]{kW}. This corresponds to the maximal expected power during O5 for a laser power of \unit[$130$]{W}. The EM and IM baffles receive \unit[$0.22$]{W} and \unit[$0.76$]{W}, respectively, when realistic mirror maps are used, as compared to \unit[$0.10$]{W} and \unit[$0.01$]{W} when we consider perfect mirrors. For this reason, all the computations of the power received by sensors presented in the remainder of this paper are done with realistic mirror maps.

\begin{table}[htp]
    \centering
    \begin{tabular}{l|l}
    \hline
    \multicolumn{2}{c}{Simulation output} \\
    \hline
    Cavity power     &  \unit[$393$]{kW}\\
    Power in EM baffle     & \unit[$2.2 \cdot 10^{-1}$]{W} ($0.6$ ppm) \\
    Power in IM baffle     & \unit[$7.6 \cdot 10^{-1}$]{W} ($1.9$ ppm) \\
    Beam width on EM     & \unit[$9.1$]{cm} \\
    Beam width on IM     & \unit[$4.9$]{cm} \\
    \end{tabular}
    \caption{Output of the simulation for the nominal configuration and realistic mirrors considered.}
    \label{tab:output_nominal}
\end{table}

% Description of sensors: area, layout...
% Plot of the layout

\begin{figure}[htp]
    \centering
    \includegraphics[width=0.8\columnwidth]{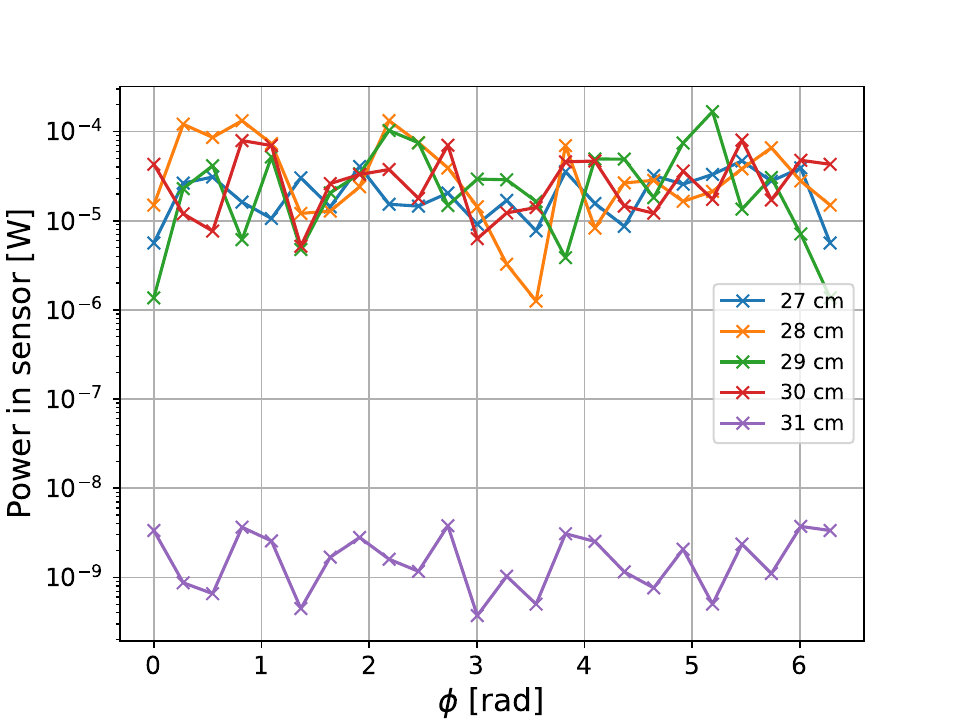}
    \caption{Power seen by each sensor in nominal configuration. Values are displayed for each ring as a function of the radial angle $\phi$.}
    \label{fig:sensors_nominal}
\end{figure}

The amount of power reaching each sensor is shown in Figure~\ref{fig:sensors_nominal}. The average value is of the order of \unit[$10^{-4}$]{W} for the first four rings, and \unit[$10^{-9}$]{W} for the fifth ring at \unit[$31$]{cm} from the center. This sharp drop is due to the cryobaffle that clips the field at a radius of \unit[$30$]{cm} close to the EM.
Because of the mirrors' roughness, the distribution of scattered light is not isotropic: in the innermost ring, the expected power in sensors varies between \unit[$5.62 \cdot 10^{-6}$]{W} and \unit[$4.69 \cdot 10^{-5}$]{W}, \newchanges{with a standard deviation of \unit[$1.18 \cdot 10^{-5}$]{W}. Since the resolution of the sensors is \unit[$6 \cdot 10^{-5}$]{W}, it is unlikely that these angular variations will be detected.}

\subsection{Misaligned cavity}

\begin{figure}[htp]
    \centering
    \includegraphics[width=0.8\columnwidth]{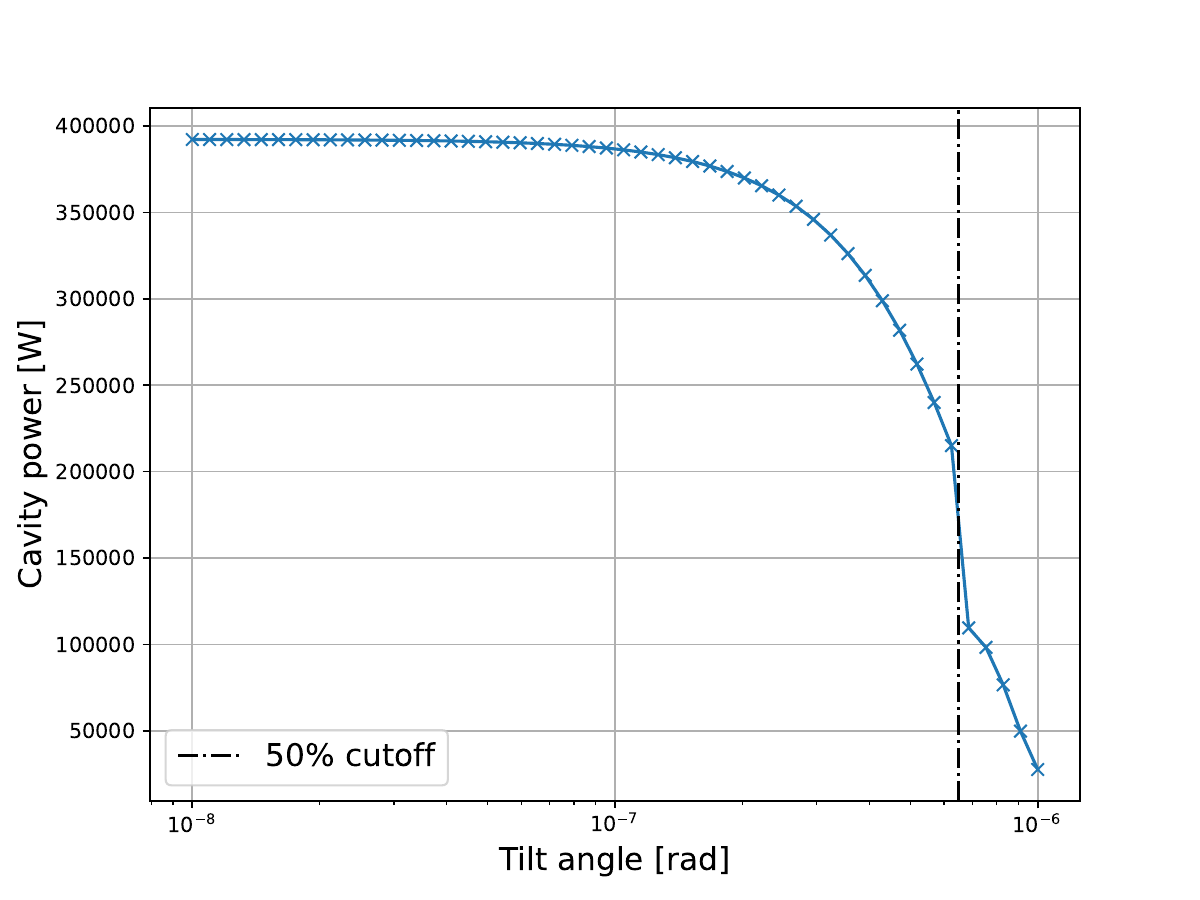}
    \caption{Total power circulating in the cavity as a function of the tilt angle applied to the EM.}
    \label{fig:P_vs_tiltX}
\end{figure}

Spurious misalignment of the interferometer's optics can induce the development of higher order modes in the laser beam and decrease the power circulating in the cavity, leading to an overall loss of sensitivity for the detector~\cite{1996PhLA..217...90B}.
One of the main interests of instrumented baffles is to be able to detect such a misalignment.
We simulate a misaligned cavity by introducing a tilt angle in the EM's surface map. This is equivalent to tilting the laser beam by the same angle in the opposite direction. As shown in Figure~\ref{fig:P_vs_tiltX}, the total power circulating in the cavity decreases as the tilt angle increases. The critical range for this value is between \unit[$10^{-7}$]{rad}, where the power starts to decrease by a few percents, to \unit[$10^{-6}$]{rad}, where the total power has decreased so much that the cavity becomes out of resonance. \revisions{At \unit[$6.1 \cdot 10^{-7}$]{rad}, $50\%$ of the power is lost.}

\begin{figure}[htp]
    \centering
    \includegraphics[width=0.45\columnwidth]{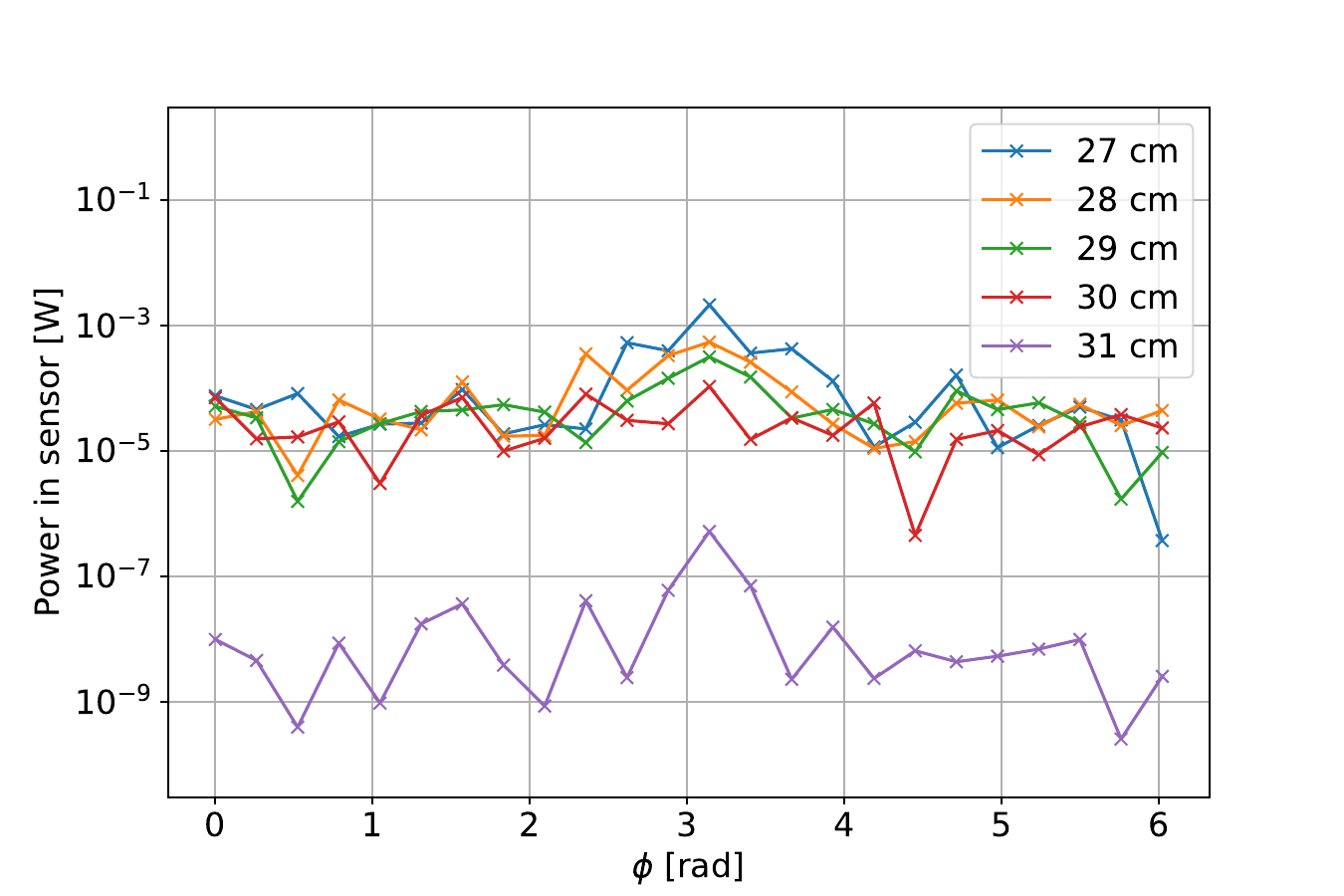}
    \includegraphics[width=0.45\columnwidth]{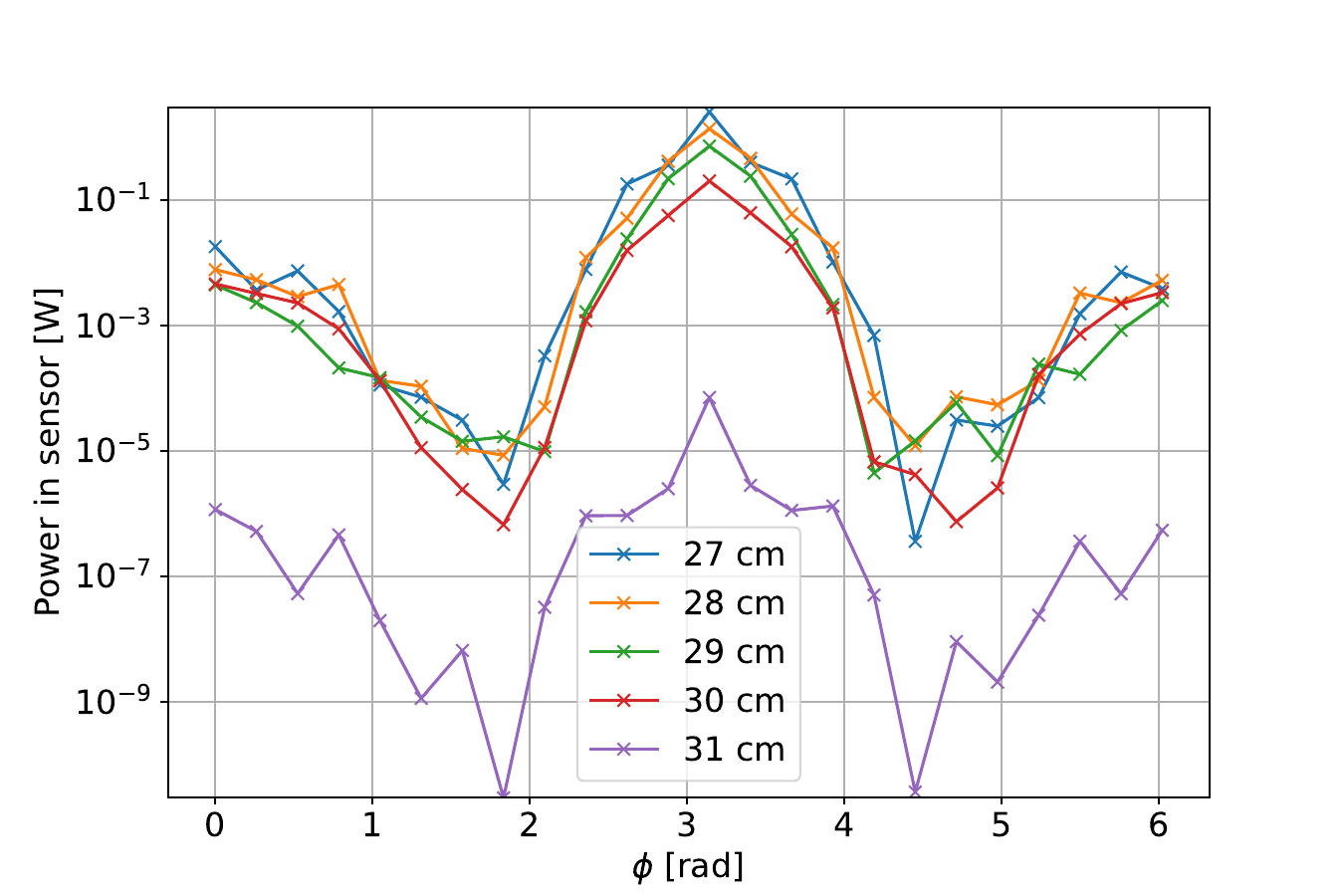}
    \caption{Power seen by each sensor when the EM is tilted by \unit[$3 \cdot 10^{-7}$]{rad} (left), and \unit[$8 \cdot 10^{-7}$]{rad} (right) along the $x$ direction.}
    \label{fig:sensors_misaligned}
\end{figure}

We illustrate both regimes in Figure~\ref{fig:sensors_misaligned} by computing the amount of power seen by each sensor for a tilt angle of \unit[$3 \cdot 10^{-7}$]{rad}, where the total power is around $90\%$ of the value expected for a perfectly aligned cavity, and for an angle of \unit[$8 \cdot 10^{-7}$]{rad}, for which this value drops to $25\%$. 

\begin{figure}[htp]
    \centering
    \includegraphics[width=0.8\columnwidth]{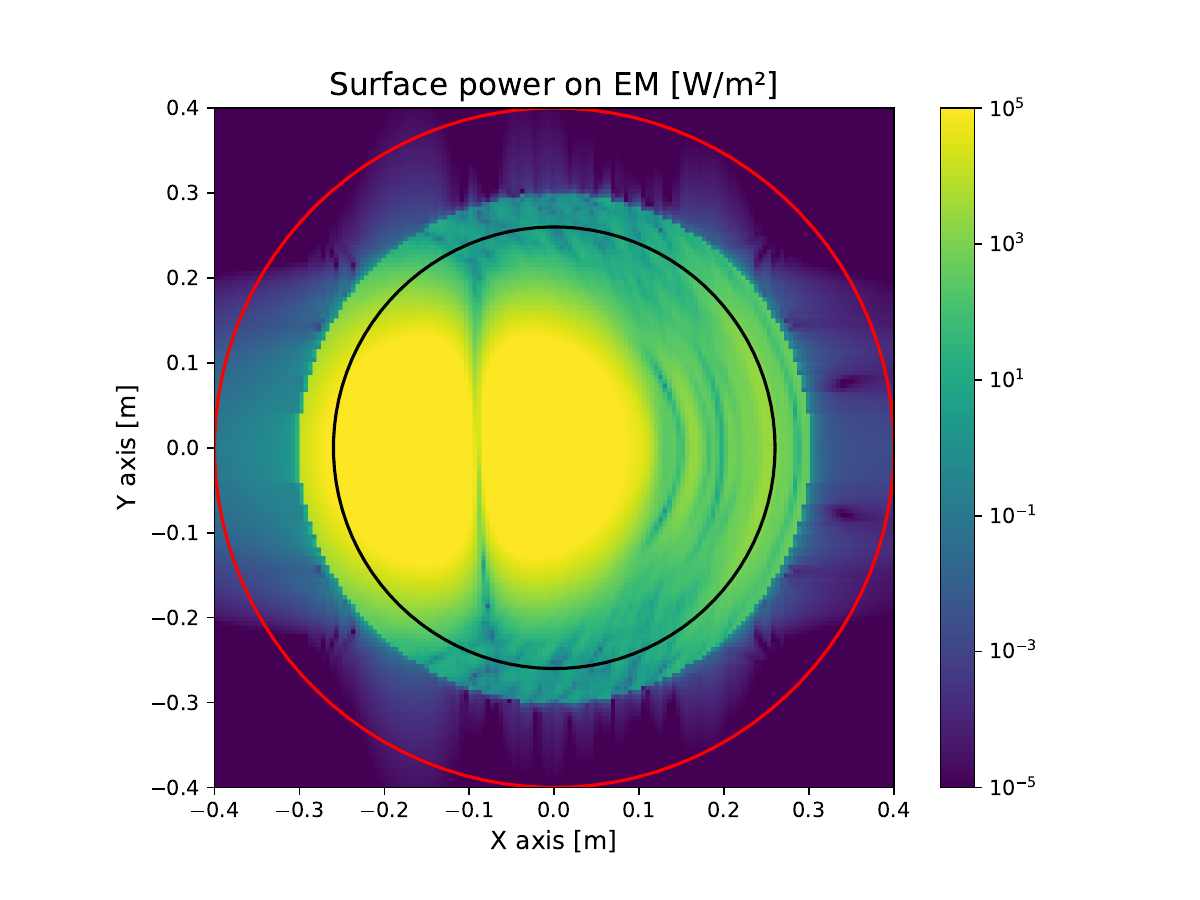}
    \caption{Surface power distribution on the EM when it is tilted by \unit[$8 \cdot 10^{-7}$]{rad} (right) along the $x$ direction.}
    \label{fig:sensors_misaligned_2d}
\end{figure}

Since the beam does not hit the center of the mirror, an excess of power is seen in the sensors located around $\phi = \pi$, the direction of misalignment. Sensors located at lower radius receive more power. In the \unit[$3 \cdot 10^{-7}$]{rad} case, a sensor at \unit[$27$]{cm} from the center of the mirror would see at most \unit[$2.1$]{mW} of power, around $50$ times more than in the nominal configuration. Therefore, such misalignment would be detectable by the sensors.
When the tilt angle is \unit[$8 \cdot 10^{-7}$]{rad}, \revisions{higher order modes of the beam start to develop. In this case, the Hermite Gaussian mode $(1,0)$ is excited, as can be seen in Fig. \ref{fig:sensors_misaligned_2d}. This explains the two dips around $\pi /2$ and $3\pi/2$ in the right panel of Fig. \ref{fig:sensors_misaligned}}.
Besides, sensors located in the first four rings would receive more than \unit[$1$]{W} of power, which would almost certainly lead them to saturate. In this case of a large misalignment, sensors located in the fourth and fifth rings could continue to provide information after sensors from the inner rings saturate. 
\revisions{As the tilt angle increases, sensors located in the direction of misalignment receive a higher fraction of the total power. However, the cavity power decreases quickly, as shown in Fig. \ref{fig:P_vs_tiltX}. We find that the value of \unit[$8 \cdot 10^{-7}$]{rad} corresponds to the maximum power a sensor would see in the case of a misaligned cavity.}

\begin{figure}[htp]
    \centering
    \includegraphics[width=0.45\columnwidth]{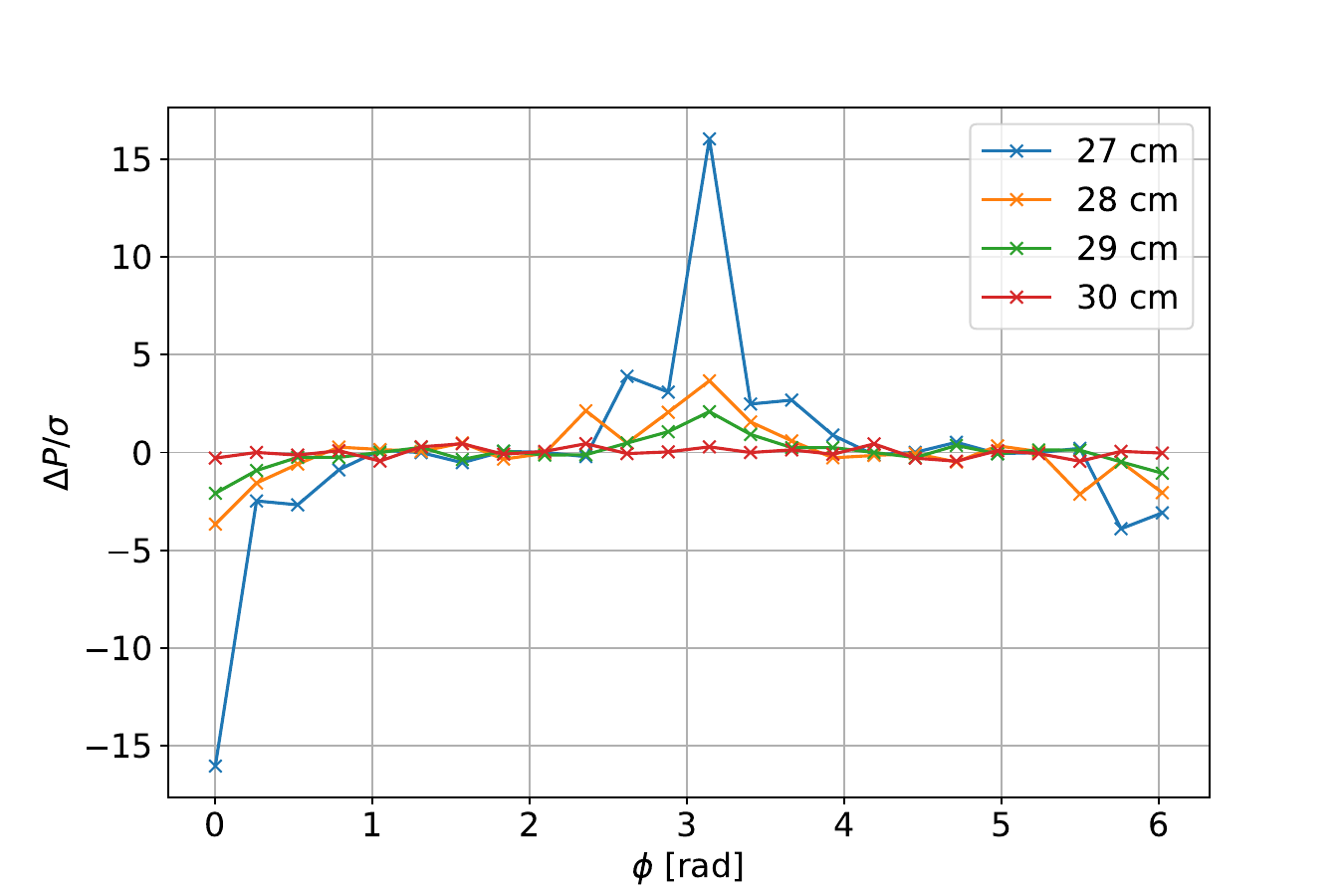}
    \includegraphics[width=0.45\columnwidth]{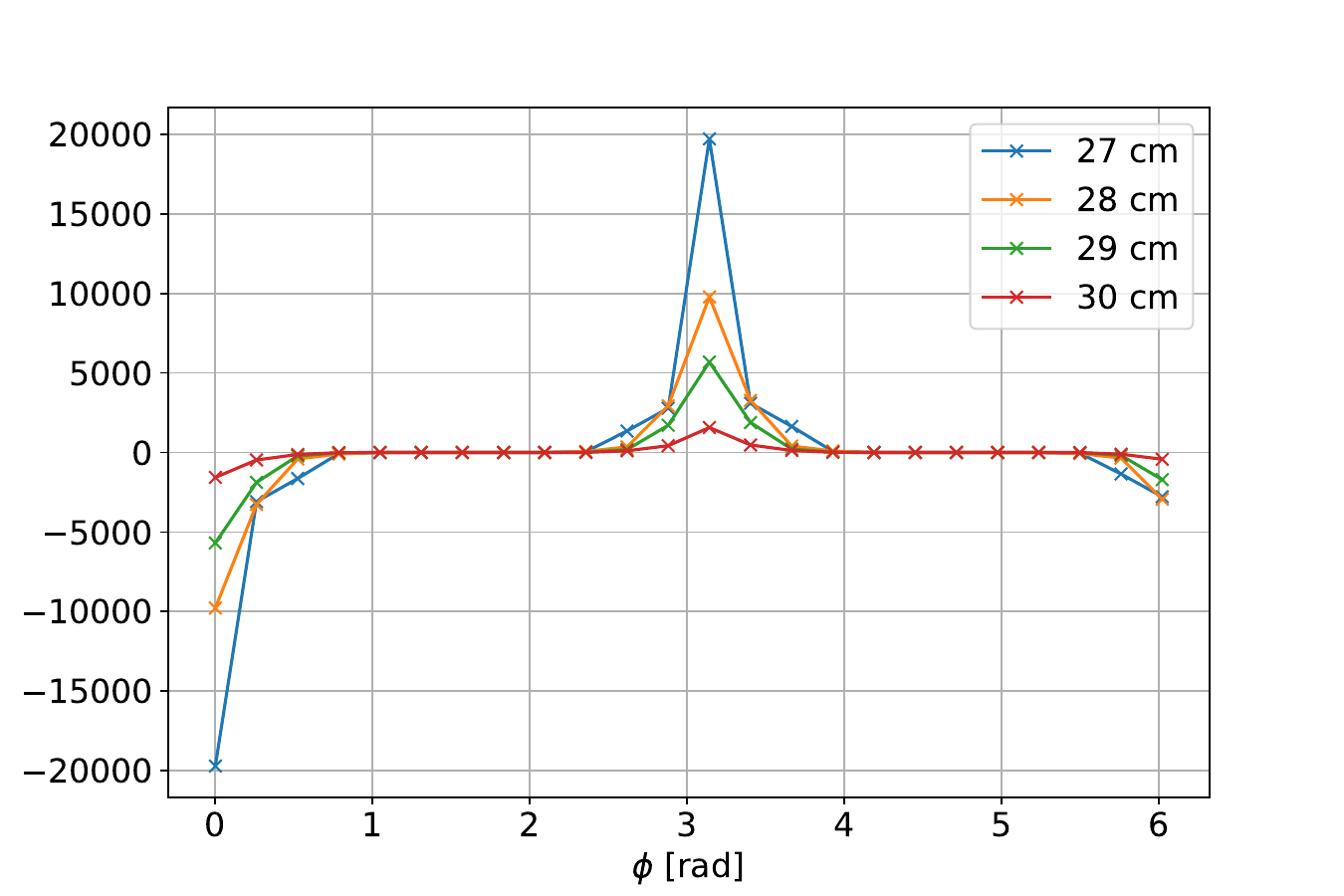}
    \caption{Distribution of the differential power in sensors $\Delta P(\phi, r) / \sigma_r$ as a function of the sensor's angular position $\phi$ and distance $r$ to the center of the mirror, in the case where the EM is tilted by \unit[$3 \cdot 10^{-7}$]{rad} (left), and \unit[$8 \cdot 10^{-7}$]{rad} (right) along the $y$ axis.}
    \label{fig:DeltaP}
\end{figure}

To show quantitatively how the sensors could be used to monitor misalignment, we define the differential power
\begin{equation}
    \Delta P(\phi, r) = P(\phi, r) - P(\phi + \pi, r),
\end{equation}
that is the difference of power in opposite sensors on the baffle. In the case of a perfectly aligned cavity, for a given ring of sensors at radius $r$, the mean value of $\Delta P(\phi, r)$ along $\phi$ is $0$ and it has a standard deviation $\sigma_r$ that comes from scattered light due to the mirrors' roughness \newchanges{and the intrinsic resolution of the sensors}. We compute signal to noise ratios $\Delta P(\phi, r) / \sigma_r$ for different values of misalignment and show their distribution in Figure~\ref{fig:DeltaP}. These distributions present a positive peak in the direction of misalignment, and a negative one in the opposite direction, which reflects the fact that the beam is off-centered with respect to the mirror. These characteristic features should be helpful to detect and correct misalignment.

%\begin{figure}[h!]

\begin{figure}[htp]
    \centering
    \includegraphics[width=0.8\columnwidth]{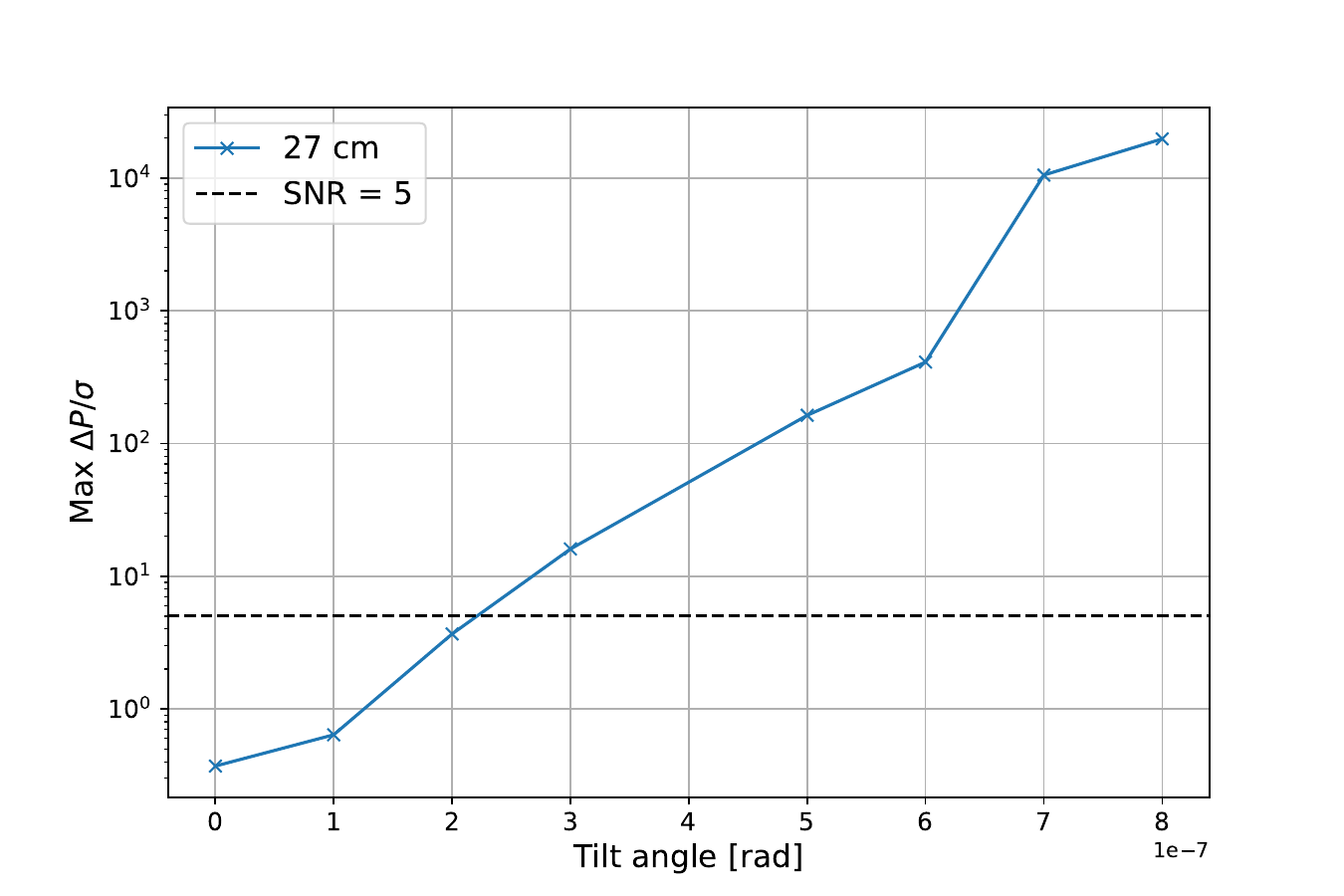}
    \caption{Maximal value of the SNR $\Delta P(\phi, r) / \sigma_r$ for the $27$ cm ring as a function of the tilt angle applied to the EM. The black dashed line represents an SNR of $5$.}
    \label{fig:maxDP_vs_tilt}
\end{figure}

To estimate the range of misalignment that the sensors would be able to monitor, we show in Figure~\ref{fig:maxDP_vs_tilt} the maximal SNR in a single sensor as a function of the tilt angle. Since sensors closer to the center of the mirror are more sensitive to misalignment, only the $27$ cm ring is represented. The excess of power starts to be significant (SNR $>5$) just above \unit[$0.2$]{$\mu$rad}, and increases exponentially with the angle until \unit[$0.6$]{$\mu$rad}. The range \unit[$0.2 - 0.6$]{$\mu$rad} corresponds to the power in the cavity decreasing from $95$\% to $50$\% of the power in the perfectly aligned configuration.

% A very rough order of magnitude is given by the angular diameter of the beam at \unit[$27$]{cm}, which is $\sim 20^{ \circ}$, corresponding to $18$ sensors.
\subsection{Point absorbers}
Point absorbers (PA) are small \revisions{, typically around \unit[$ 0.1$]{mm}}, highly absorbing areas located on the surface of the mirrors, that generate non-uniform absorption. Several of them have been detected in Virgo and the two LIGO detectors during O3~\cite{2021ApOpt..60.4047B, Cifaldi:2022bjq}. 
These PAs are harmful to the sensitivity of the detector, as they scatter power from the fundamental resonating mode of the beam into higher order modes, decreasing the overall circulating power.  
In order to detect these defects, one needs to physically inspect the mirrors, which is impractical, and cannot be done during observing runs. That is why we are interested in assessing the ability of instrumented baffles to monitor them.

A PA is simulated in SIS by adding a local surface deformation in the mirror map. It is characterized by its location on the mirror and its total absorbing power. During O3, PAs have been found at distances between $0.1$ and $5.4$~cm from the center of the mirrors, and their absorbing power estimated around $10$~mW \cite{Cifaldi:2022bjq}. For this reason, we set an absorbing power of $10$ mW for the PAs studied here. 

\begin{figure}[htp]
    \centering
    \includegraphics[width=0.8\columnwidth]{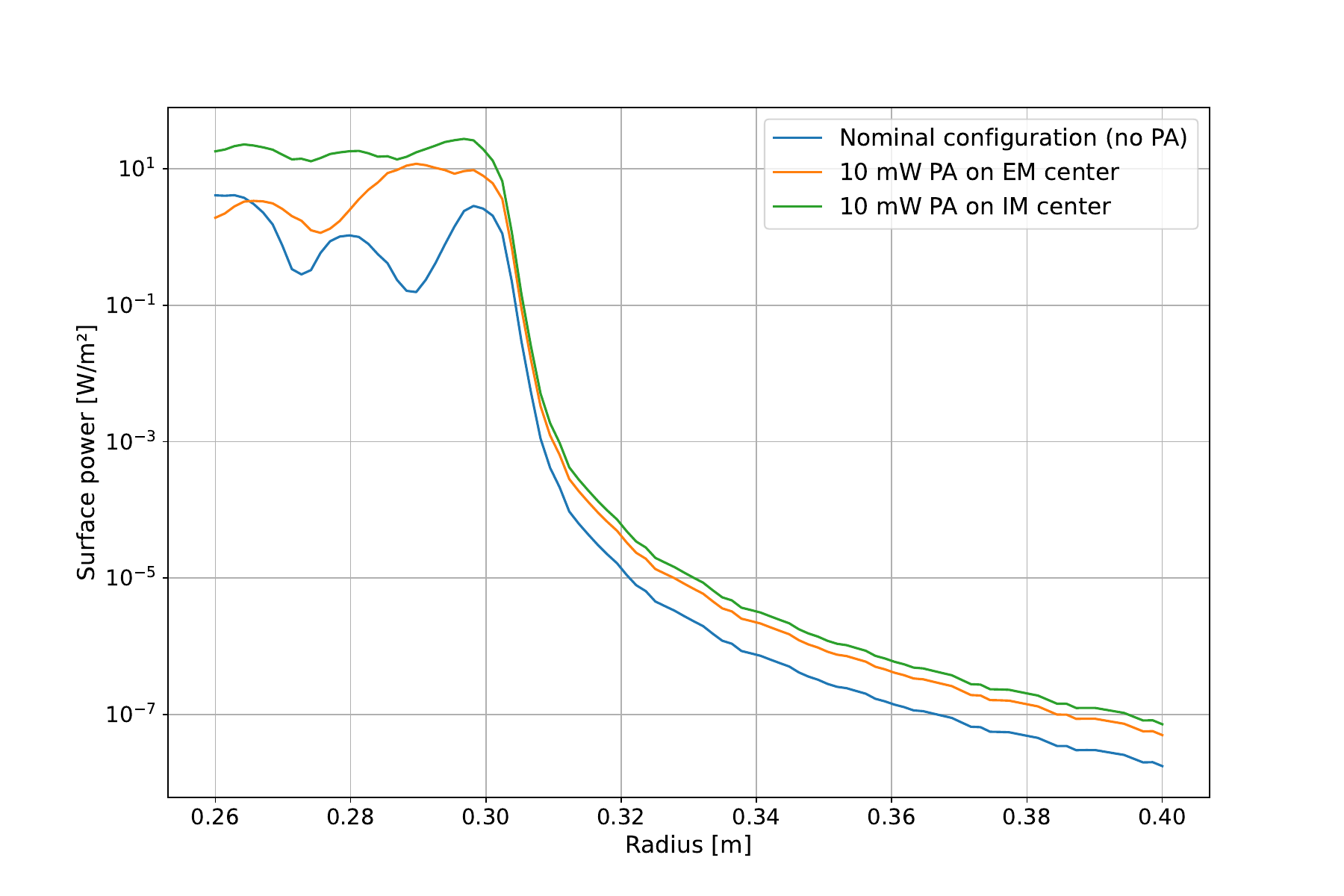}
    \caption{Radial distribution of power on the EM baffle when a $10$ mW PA is placed at the center of the EM (orange), and the IM (green). The nominal case where no PA is added is represented in blue.}
    \label{fig:PA_radial}
\end{figure}

\begin{figure}[htp]
    \centering
    \includegraphics[width=0.45\columnwidth]{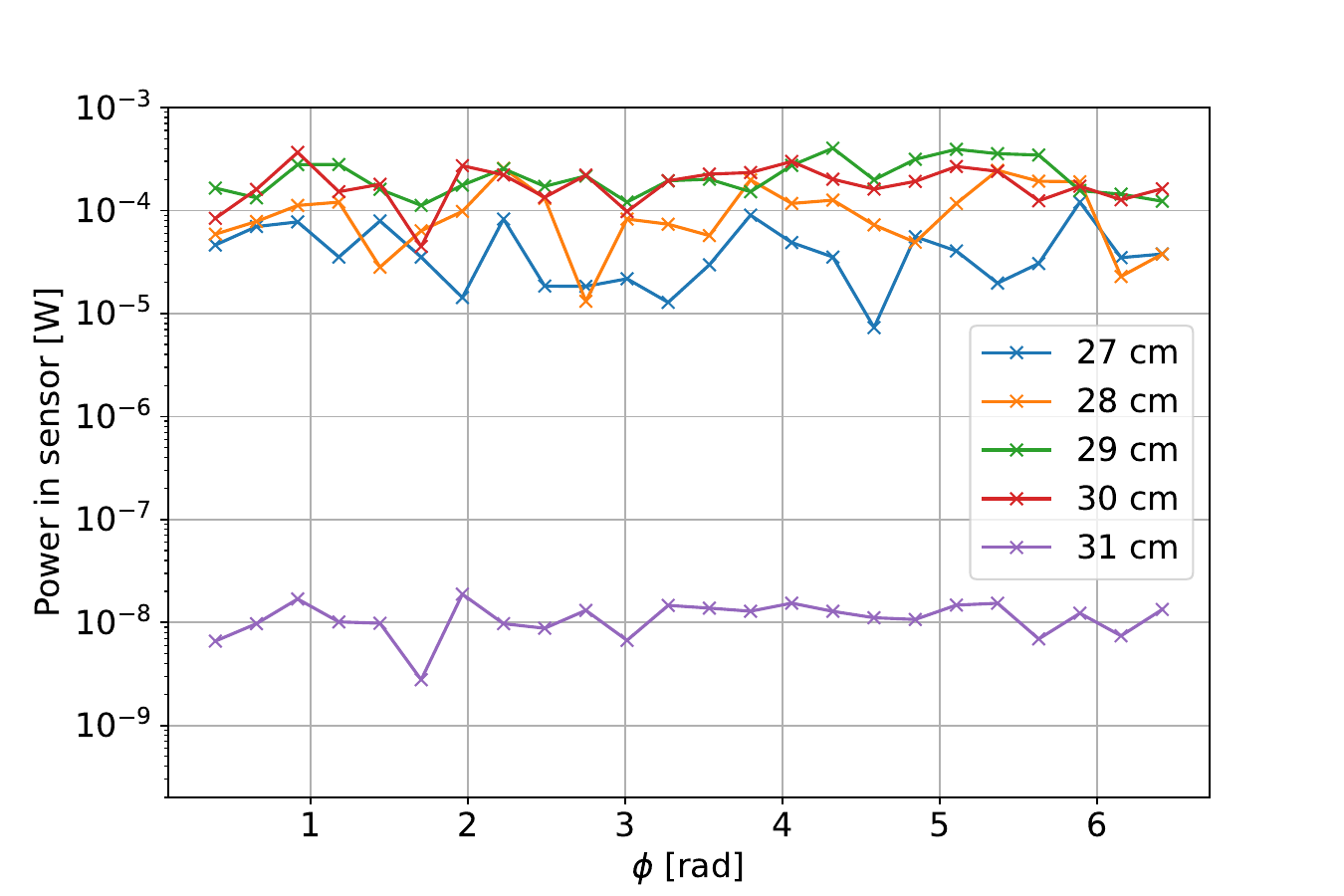}
    \includegraphics[width=0.45\columnwidth]{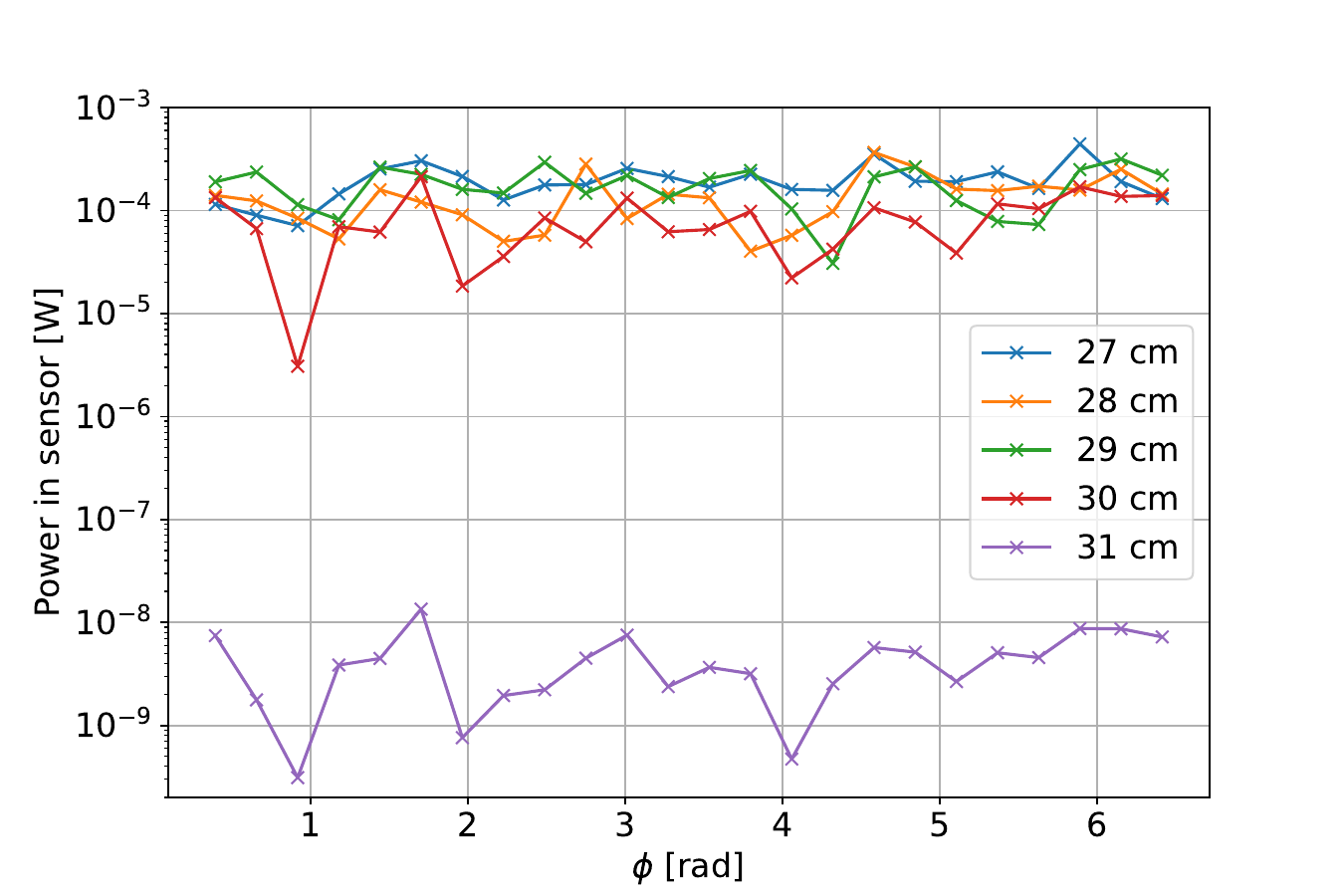}

    \caption{\revisions{Power seen by each sensor when a PA is placed at the center of the IM (left), and the EM (right). Each color represents a different ring of sensors.}
}
    \label{fig:PA_sensors_center}
\end{figure}

Figure~\ref{fig:PA_radial} shows the radial distribution of power on the EM baffle when a $10$ mW PA is placed at the center of EM, and the IM, respectively. In both cases, an excess of power is present as compared to the nominal configuration. When the PA is located on the IM, the surface power is globally constant from $26$ to $30$~cm, and more than $10$ times higher than in the absence of a PA.
When a PA is placed on the center of the EM, surface power tends to increase with the radius, reaching a plateau at $29$-$30$~cm.
The corresponding power seen by each sensor is shown in Figure~\ref{fig:PA_sensors_center}. Hence, the signature of a PA would be clearly visible as an excess of power in sensors in the first four rings. Contrary to the case of a misalignment, the power does not decrease with radius. \newchanges{Besides, the fact that PAs could be present on both the EM and IM motivates the instrumentation of the baffles around the IM as well.}

\begin{figure}[htp]
    \centering
    \includegraphics[width=0.45\columnwidth]{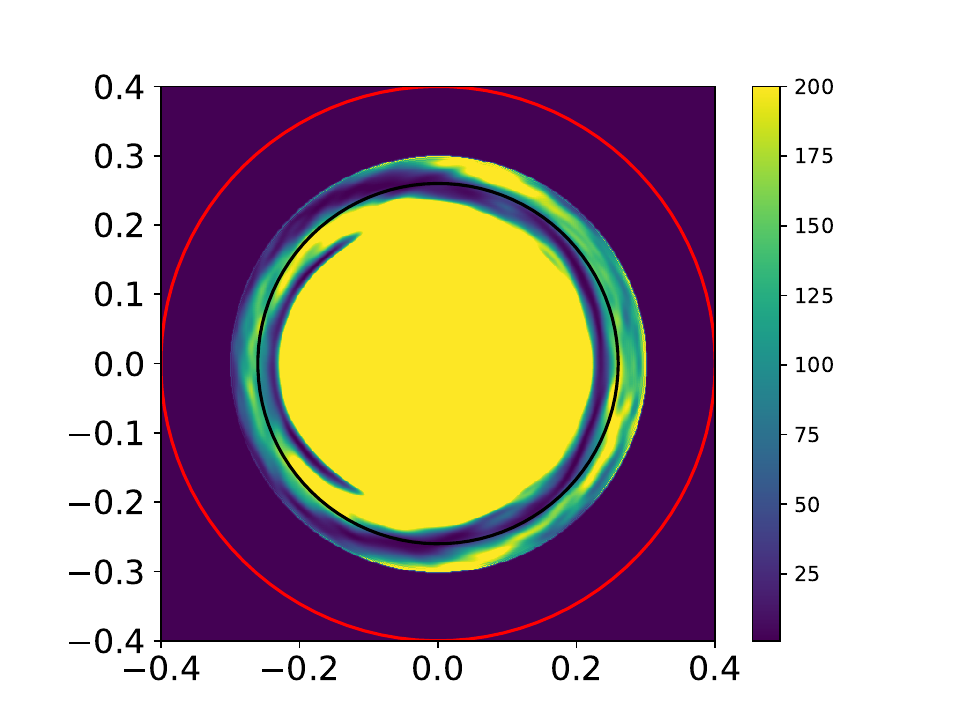}
    \includegraphics[width=0.45\columnwidth]{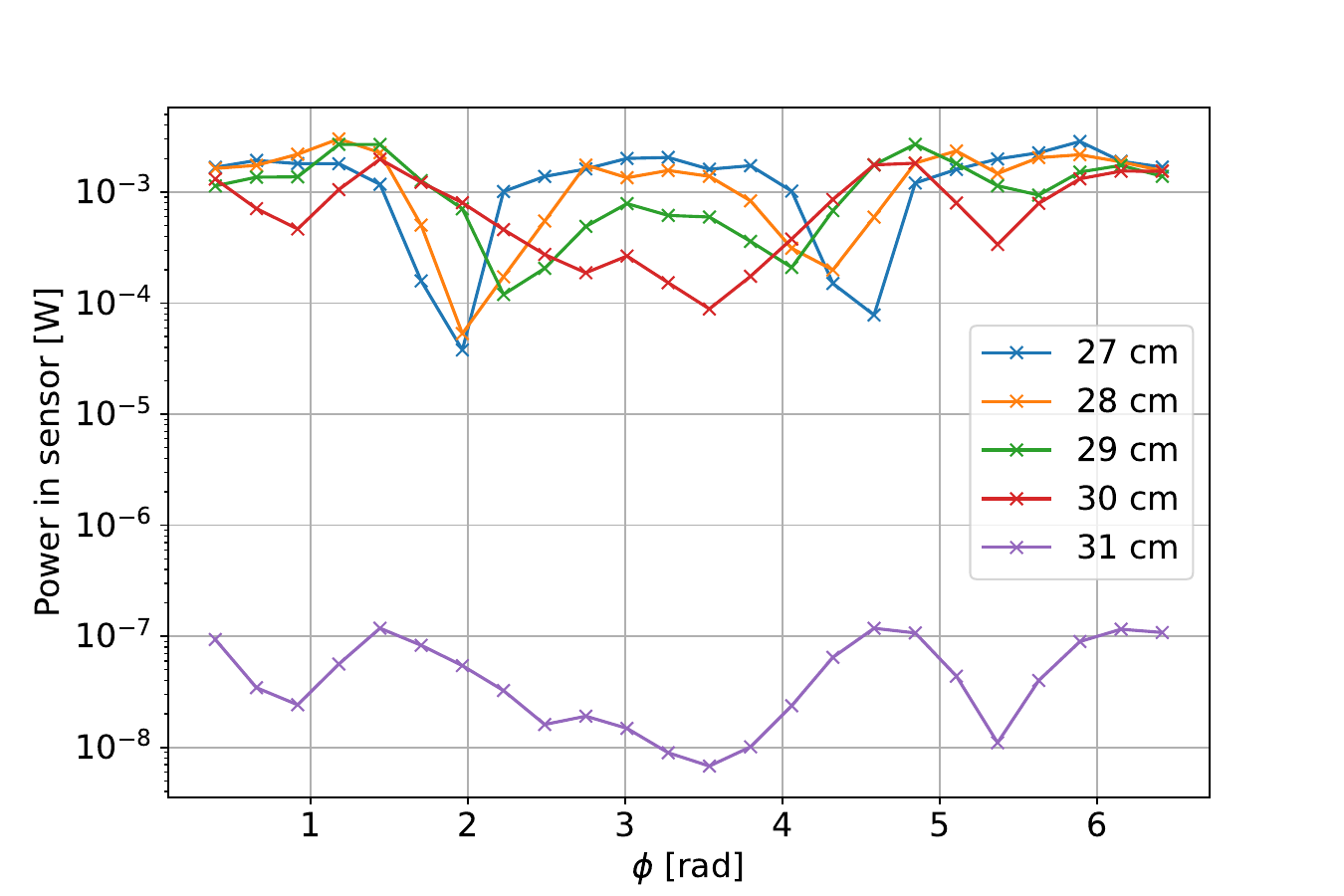}
    \caption{Surface power distribution on the EM and baffle (left) and distribution of power in sensors (right) when a \unit[$50$]{mW} PA is placed at \unit[$5$]{cm} to the right of the center of the EM.}
    \label{fig:2D_PA_offcenter}
\end{figure}

To study the case where a PA is away from the center of the mirror, we place a \unit[$50$]{mW} PA at \unit[$5$]{cm} to the right of the center of the EM. The resulting surface power distribution and power in sensors are shown in Figure~\ref{fig:2D_PA_offcenter}. As for the centered PA, there is a global excess of power in all rings that indicates the presence of a PA. The angular distribution of power is not isotropic due to the presence of Airy rings centered at the position of the PA. Nevertheless, the coverage provided by the sensors is not sufficient to confidently infer the precise position of the PA.

Finally,
Table~\ref{tab:max_power_in_sensor} presents the maximal values of power reaching
an individual sensor in each ring and for every configuration. \revisions{A Laser Induced Damage Threshold test has been performed to check the resistance of the sensors to high laser power. It has shown that for a surface power up to \unit[$50$]{$\textrm{W/cm}^2$}, sensors saturate but are not damaged.}

\begin{table}[htp]
    \centering
    \begin{tabular}{c|c|c|c|c}
        \hline 
         & \multicolumn{4}{c}{Maximal power in one sensor [W]}  \\
         Radius [cm] & Nominal & \unit[$0.3$]{$\mu \rm{rad}$ tilt} & \unit[$0.8$]{$\mu \rm{rad}$ tilt} & \unit[$10$]{mW} point absorber \\
        \hline
         $27$& $4.7 \cdot 10^{-5}$ & $2.1 \cdot 10^{-3}$ & $2.6$ & $1.8 \cdot 10^{-4}$ \\
         $28$& $1.3 \cdot 10^{-4}$ & $6.0 \cdot 10^{-4}$ & $1.4$ & $3.2 \cdot 10^{-4}$\\
         $29$& $1.7 \cdot 10^{-4}$ & $4.0 \cdot 10^{-4}$ & $7.3 \cdot 10^{-1}$ & $6.5 \cdot 10^{-4}$\\
         $30$& $8.0 \cdot 10^{-5}$ & $1.9 \cdot 10^{-4}$ & $2.0 \cdot 10^{-1}$ & $4.2 \cdot 10^{-4}$ \\
         $31$& $3.8 \cdot 10^{-9}$ & $5.2 \cdot10^{-7}$ & $7.1 \cdot 10^{-5}$ & $2.5 \cdot 10^{-8}$\\

    \end{tabular}
    \caption{Maximal power seen by a single sensor in each ring for different configurations.}
    \label{tab:max_power_in_sensor}
\end{table}

%\section{Projected impact on Virgo sensitivity}
%\label{sec:finesse}
\section{Conclusion}

The optical simulations presented here are intended to estimate the amount of light that will be seen by sensors installed on the EM baffle in different configurations, in order to provide constraints on the sensitivity and dynamic range of sensors, and to demonstrate the ability of this instrument to detect defects in the cavity.

We have shown that the sensors will receive an \newchanges{order of magnitude} of \unit[$10^{-4}$]{W} in nominal conditions when the circulating power in the cavity is \unit[$380$]{kW}, with a variability of approximately $100\%$ due to the non-isotropic scattering from the mirrors' surface.
Sensors will be able to detect the anti-symmetric difference of power due to a misalignment of the cavity when the deviation of the beam is larger than \unit[$0.2$]{$\mu$rad}. This should prove particularly helpful for pre-alignment operations.
\newchanges{Sensors will also provide meaningful information about the presence of defects on the mirrors' surface.} Simulations show that sensors can confidently detect the signature of a \unit[$10$]{mW} PA near the center of the EM or the IM. \newchanges{While inferring the precise number and positions of PAs remains a difficult goal to achieve, 
instrumented baffles will allow to monitor potential degradation of the mirrors in real time.}

% \todo{Conclusion on backscaterring.}

\bigskip\noindent\textit{Acknowledgments} ---
%LIGO and Virgo
The authors gratefully acknowledge the European Gravitational Observatory
(EGO) and the Virgo Collaboration for providing access to the facilities. The
LIGO Observatories were constructed by the California Institute of Technology and Massachusetts Institute of Technology with funding from the National
Science Foundation under cooperative agreement PHY-9210038. The LIGO
Laboratory operates under cooperative agreement PHY-1764464.
% ProBIST grant acknowledgment
This project has received funding from the European Union’s Horizon 2020 research and
innovation programme under the Marie Skłodowska-Curie grant agreement No. 754510.
% IFAE aknowledgment
This work is partially supported by the Spanish MCIN/AEI/10.13039/501100011033 under the Grants No. SEV-2016-0588, No. PGC2018-101858-B-I00,   and No. PID2020-113701GB-I00, some of which include ERDF  funds  from  the  European  Union, and by the MICINN with funding from the European Union NextGenerationEU (PRTR-C17.I1) and by the Generalitat de Catalunya. IFAE  is  partially funded by the CERCA program of the Generalitat de Catalunya. 
MAC is supported by the 2022 FI-00335 grant.
% ARR is supported in part by the Strategic Research Program “High-Energy Physics” of the Research Council of the Vrije Universiteit Brussel and by the iBOF “Unlocking the Dark Universe with Gravitational Wave Observations: from Quantum Optics to Quantum Gravity” of the Vlaamse Interuniversitaire Raad and by the FWO IRI grant I002123N “Essential Technologies for the Einstein Telescope”.
\section*{References}

\bibliographystyle{unsrt}
\bibliography{biblio.bib}

\end{document}